\documentclass[twocolumn, apj]{emulateapj}
\usepackage{natbib, amsmath}
\newcommand{\me}{\mathrm{e}} 

\newcommand{\dif}{\mathrm{d}}
\newcommand{\degree}{\ensuremath{^\circ}}
\def\la{\langle}
\def\ra{\rangle}
\def\ch{{\em Chandra} }
\def\xmm{{\em XMM-Newton} }

\def\fspitz           {f_{\textrm{Sp}}}
\def\om{\omega}
\def\omt{\tilde \omega}
\def\omc{\omega_{\rm cond}}

\newcommand       \be           {\begin{equation}}
\newcommand       \ee           {\end{equation}}

\begin{document}
\title{Anisotropic Thermal Conduction and the Cooling Flow Problem in Galaxy Clusters}
\author{Ian J. Parrish\altaffilmark{1,2}, Eliot Quataert\altaffilmark{1}, \& Prateek Sharma\altaffilmark{1,2}}
\altaffiltext{1}{Astronomy Department \& Theoretical Astrophysics Center, 601 Campbell Hall, University of California, Berkeley, CA 94720; iparrish@astro.berkeley.edu}
\altaffiltext{2}{Chandra/Einstein Fellow}
\shorttitle{Thermal Conduction and the Cooling Flow Problem}
\shortauthors{Parrish, et al}

\begin{abstract}
We examine the long-standing cooling flow problem in galaxy clusters with 3D MHD simulations of isolated clusters including radiative cooling and anisotropic thermal conduction along magnetic field lines.  The central regions of the intracluster medium (ICM) can have cooling timescales of $\sim200$ Myr or shorter---in order to prevent a cooling catastrophe the ICM must be heated by some mechanism such as AGN feedback or thermal conduction from the thermal reservoir at large radii.  The cores of galaxy clusters are linearly unstable to the heat-flux-driven buoyancy instability (HBI), which significantly changes the thermodynamics of the cluster core.  The HBI is a convective, buoyancy-driven instability that rearranges the magnetic field to be preferentially perpendicular to the temperature gradient.  For a wide range of parameters, our simulations demonstrate that in the presence of the HBI, the effective radial thermal conductivity is reduced to $\lesssim 10$\% of the full Spitzer conductivity.  With this suppression of conductive heating, the cooling catastrophe occurs on a timescale comparable to the central cooling time of the cluster.  Thermal conduction alone is thus unlikely to stabilize clusters with low central entropies and short central cooling timescales.  High central entropy clusters have sufficiently long cooling times that conduction can help stave off the cooling catastrophe for cosmologically interesting timescales.  
\end{abstract}
\keywords{convection---galaxies: clusters: general---instabilities---MHD---plasmas---X-rays: galaxies: clusters}
\section{Introduction}\label{sec:intro}
X-ray observations of the intracluster medium (ICM) of relaxed galaxy clusters show a centrally peaked surface brightness distribution.  The observed temperatures and densities are high enough that the plasma can have a cooling time much less than 500 Myr \citep{sarazin86}.  The standard isobaric cooling flow model predicts mass dropping out of the X-ray emitting ICM at rates in excess of 100--$500\,M_{\odot}\,\textrm{yr}^{-1}$.  However, X-ray spectroscopic observations with \ch and \xmm have ruled out classical cooling flows of material below 1 keV as it would copiously emit lines such as Fe XVII that are not observed \citep{pf06}.  Therefore, a mechanism is required to heat the ICM to avert this cooling catastrophe.  

The plasma in the ICM has temperatures of 1--15 keV and number densities of $10^{-3}$--$10^{-1}$ cm$^{-3}$.  The magnetic field in the ICM is estimated to be in the range of 0.1--10 $\mu$G depending on where the measurement is made \citep{ct02}.  Under these conditions, the Coulomb mean free path is many orders of magnitude larger than the gyroradius; e.g., at $T=3$ keV, $n_e = 10^{-2}\,\textrm{cm}^{-3}$, and $B=1\,\mu$G, the mean free path is $\lambda_{\textrm{mfp}}\approx 0.3$ kpc, while the electron gyroradius is $\rho_e\approx 10^8$ cm. The mean free path is, however, smaller than the temperature gradient scale length. As a result, a fluid description of the ICM plasma, e.g. MHD, is appropriate, but the effects of anisotropic heat and momentum transport must also be included. The Braginskii-MHD equations \citep{brag65} are a modification of the standard MHD equations to include anisotropic transport due to the magnetic field.  

AGN feedback and conduction from the thermal bath at large radii are two of the most often discussed mechanisms for heating cool cluster cores.  This work shall only briefly consider the former.  The latter, thermal conduction, has been studied by many authors \citep[e.g., ][]{bc81, nm01, zn03, gor08}.  Because of uncertainties associated with the magnetic geometry, the heat flux is often parameterized as an effective thermal conductivity given as a fraction, $f_{\textrm{Sp}}$, of the ideal (field-free) Spitzer heat flux.  Previous work, however, has not considered the dynamic consequences of anisotropic conduction.  The plasma in galaxy clusters is unstable to two different convective instabilities driven by anisotropic thermal conduction along magnetic field lines.  The first, the magnetothermal instability, or MTI \citep{bal00,psl08}, occurs when the temperature gradient and gravity are in the same direction, as is true at large radii in galaxy clusters.  
Well inside the cooling core, the heat-flux-driven-buoyancy instability, or HBI, occurs where the temperature gradient is in the opposite direction of gravity \citep{quat08,pq08}.  The HBI occurs in the cooling core of the ICM where the temperature increases outward.  Nonlinear simulations of the MTI and HBI have shown that they significantly modify the thermal conductivity, because they saturate by rearranging the magnetic field geometry.  Thus, determining the effective conductivity in the ICM requires considering the back-reaction of the anisotropic heat flux on the magnetic field geometry.

In this work, we examine the stability of the cooling cores of galaxy clusters using three-dimensional MHD simulations including anisotropic thermal conduction and cooling.  In particular we assess the interplay between cooling and the HBI.  This paper is organized as follows.  In \S\ref{sec:HBIphys} we summarize the physics of the HBI and the thermal instability and their nonlinear saturation in local calculations.  In \S\ref{sec:method} we then describe the equations of Braginskii-MHD and the numerical tools we utilize to solve them.  \S\ref{sec:fiducial} presents our fiducial cluster model (based on Abell 2199) in detail and examines its evolution over cosmic time.  In \S\ref{sec:paramvar}, we examine a variety of variations of our fiducial model including cluster halo parameters, magnetic field strength and geometry, and the central cluster entropy.  We also show a few experiments with a simple AGN heating model (\S\ref{sec:heating}).  Finally, in \S\ref{sec:conclusions} we discuss the implications of this work for the cooling flow problem and highlight some of our plans for future work.

In parallel to our work described here, \citet{bog09} has conducted a similar study using similar numerical methods.  Their simulations start with different initial conditions and cover a different part of parameter space but reach broadly similar conclusions.  
\section{Physics of the HBI and Cooling}\label{sec:HBIphys}
The linear physics of the HBI and its nonlinear evolution are outlined in \citet{quat08} and \citet{pq08}, respectively.  We briefly review them here for clarity.   The heat-flux-driven buoyancy instability is a convective instability driven by a background heat flux with the temperature gradient as the source of free energy.  In contrast, the entropy gradient drives the more familiar Schwarzschild convection.  For an arbitrarily oriented magnetic field, the dispersion relation of the HBI is \citep{quat08}:
\begin{eqnarray} \label{eqn:DR} &0& =  \om \omt^2
 + i \omc \omt^2 - N^2 \omega {k_\perp^2 \over k^2} \nonumber \\ &-& i
 \omc g \left({d\ln T \over dz}\right) \left[(1 - 2b_z^2){k_\perp^2
 \over k^2} + {2b_xb_zk_xk_z \over k^2}\right], 
\end{eqnarray}
where $N$ is the Brunt-V\"ais\"al\"a frequency,
\begin{equation}
N^{2} = -\frac{1}{\gamma \rho}\frac{\partial P}{\partial z}\frac{\partial \ln S}{\partial z},
\label{eqn:brunt-vaisala}
\end{equation}
and
\begin{equation}
\omt^2 = \om^2 - \left(\boldsymbol{k}\cdot\boldsymbol{v_{A}}\right)^2,
\label{eqn:omegatilde}
\end{equation}
where $\boldsymbol{v_A} = \boldsymbol{B}/(4\pi\rho)^{1/2}$ is the Alfv\'{e}n speed, and 
\begin{equation}
\om_{\textrm{cond}} = \frac{2}{5}\chi\left(\boldsymbol{\hat{b}}\cdot\boldsymbol{k}\right)^2
\label{eqn:omegacond}
\end{equation}
is the frequency for conduction to act on a given scale, with $\boldsymbol{\hat{b}}$ the unit vector directed along the magnetic field and $\chi$ the thermal diffusivity\footnote{The literature is not consistent regarding the use of $\chi$ and $\kappa$.  We will use $\chi$ to represent a true diffusion coefficient and $\kappa$ to represent a conductivity in units erg cm$^{-1}$ s$^{-1}$ K$^{-1}$.} in units of cm$^2$s$^{-1}$.  This dispersion relation is written without loss of generality for a geometry in which gravity and the initial atmospheric gradients are in the $\boldsymbol{\hat{z}}$-direction and the initial magnetic field lies in the $\boldsymbol{\hat{x}}$--$\boldsymbol{\hat{z}}$ plane. 

For a weak magnetic field, equation (\ref{eqn:DR}) has unstable solutions for either sign of the temperature gradient.  The case of $\dif T/\dif z >0$ corresponds to the HBI.  
For a weak, initially vertical magnetic field ($b_z =1$, $b_x=0$), the growth rate of the HBI is given by
\begin{equation}
\om^2 \approx -g\left(\frac{\dif \ln T}{\dif z}\right)\frac{k_{\perp}^2}{k^2},
\label{eqn:hbigr}
\end{equation}
where $k_\perp$ is with respect to gravity.  Qualitatively, one can picture the HBI as being driven by regions of converging and diverging perturbed magnetic field lines.  Regions of converging magnetic field are conductively heated and become buoyant.  In local simulations, the HBI generates MHD turbulence and a modest magnetic dynamo that amplifies the field.  The most prominent method by which the HBI saturates is via a significant reorientation of the magnetic field geometry.  The HBI takes a largely vertical field and reorients it to become largely horizontal.  This fact is crucial for cluster cores since this reorientation of the magnetic field can significantly reduce the heat transport across an HBI unstable region.  

In addition to driving the HBI, thermal conduction also has a significant impact on thermal instability as originally shown in \citet{field65}.  The Field criterion states that wavelengths longer than
\be
\lambda_F = \left[\frac{T\kappa}{n_e n_p \Lambda(T)}\right]^{1/2},
\label{eqn:field}
\ee
 are thermally unstable, where $\Lambda(T)$ is the cooling function discussed later.  For modes with wavelengths shorter than $\lambda_F$, the conduction time is shorter than the cooling time, and local perturbations are stabilized.  In the ICM, the plasma is often locally stable to thermal instability, but unstable to global modes even in the presence of conduction \citep{nk03}. Equation (\ref{eqn:field}) was derived under the assumption of isotropic conduction.  The results are similar for anisotropic conduction, except that conduction can only stabilize perturbations with short wavelengths along the magnetic field.   

An interesting way to examine the physics of cooling in clusters comes from the recent work of \citet{voit08}, who examined the role of the central entropy of the ICM as an indicator of feedback and star formation in galaxy clusters.  The entropy is defined as $K = k_BTn_e^{-2/3}$.   In \citet{voit08}, the entropy profile for a cluster is fit using
\be
K(r) = K_0 + K_{100}\left(\frac{r}{100\,\textrm{kpc}}\right)^{\alpha},
\label{eqn:entropy}
\ee
where $K_0$ is approximately the central entropy, $K_{100}$ is the power law normalization and $\alpha > 0$ is the power law exponent.  They find that low entropy clusters, those with $K_0 \le 30$ keV cm$^{2}$, have stronger H$\alpha$ emission, star formation indicators, and AGN activity than higher entropy clusters, $K_0 \ge 30$ keV cm$^{2}$ \citep{cdv08}.  As a matter of terminology, clusters with an inwardly decreasing temperature are referred to as cool-core or relaxed clusters.  Clusters with an isothermal or inwardly increasing temperature profile are referred to as non cool-core clusters.  

These observational results can be qualitatively understood in light of the Field criterion (eq. [\ref{eqn:field}]).   When cooling is pure Bremsstrahlung, the Field length becomes a function of entropy, scaling as $\lambda_F \propto K_0^{3/2} f_{\textrm{Sp}}^{1/2}$.  Thus, cooling can take place on small length scales when the entropy is low.  The HBI decreases $f_\textrm{Sp}$, making smaller wavelengths unstable to cooling.  Fully non-linear simulations, such as those presented here, are needed to understand the combined dynamics of cooling and the HBI.   
\newpage
\section{Method}\label{sec:method}
\subsection{Equations of MHD with Anisotropic Heat Conduction} \label{subsec:method:MHDeqn}
We solve the usual equations of ideal MHD with the addition of a vector heat flux, \boldmath$Q$\unboldmath, and a gravitational acceleration $\textbf{\em g} = -\boldsymbol{\nabla}\Phi$:
\begin{equation}
\frac{\partial \rho}{\partial t} + \boldsymbol{\nabla}\cdot\left(\rho \boldsymbol{ v}\right) = 0,
\label{eqn:MHD_continuity}
\end{equation}
\begin{equation}
\frac{\partial(\rho\boldsymbol{v})}{\partial t} + \boldsymbol{\nabla}\cdot\left[\rho\boldsymbol{vv}+\left(p+\frac{B^{2}}{8\pi}\right)\boldsymbol{I} -\frac{\boldsymbol{BB}}{4\pi}\right] + \rho\boldsymbol{\nabla}\Phi=0,
\label{eqn:MHD_momentum}
\end{equation}
\begin{eqnarray}
\label{eqn:MHD_energy}
\frac{\partial E}{\partial t} &+& \boldsymbol{\nabla}\cdot\left[\boldsymbol{v}\left(E+p+\frac{B^{2}}{8\pi}\right) - \frac{\boldsymbol{B}\left(\boldsymbol{B}\cdot\boldsymbol{v}\right)}{4\pi}\right] \\ \nonumber
&+&\boldsymbol{\nabla}\cdot\boldsymbol{Q} +\rho\boldsymbol{\nabla}\Phi\cdot\boldsymbol{v}
=\mathcal{H} - \mathcal{L},
\end{eqnarray}
\begin{equation}
\frac{\partial\boldsymbol{B}}{\partial t} - \boldsymbol{\nabla}\times\left(\boldsymbol{v}\times\boldsymbol{B}\right)=0,
\label{eqn:MHD_induction}
\end{equation}
where the symbols have their usual meaning. The total energy $E$ is given by
\begin{equation}
E=\epsilon+\rho\frac{\boldsymbol{v}\cdot\boldsymbol{v}}{2} + \frac{\boldsymbol{B}\cdot\boldsymbol{B}}{8\pi},
\label{eqn:MHD_Edef}
\end{equation}
and the internal energy, $\epsilon=p/(\gamma-1)$.  We assume
$\gamma=5/3$ throughout.

The anisotropic heat flux is given by
\begin{equation}
\boldsymbol{Q} = - n k_B \chi_{C}(T, n) \boldsymbol{\hat{b}\hat{b}}\cdot\boldsymbol{\nabla}T,
\label{eqn:coulombic}
\end{equation}
where the thermal diffusivity is given by the Spitzer value \citep{spitz62} and $\boldsymbol{\hat{b}}$ is a unit vector in the direction of the magnetic field.  The Spitzer thermal diffusivity is given by
\begin{equation}
\chi_{C}(T,n) = 8\times 10^{31}\left(\frac{T}{10\,\mbox{keV}}\right)^{5/2}
\left(\frac{n}{5\times 10^{-3}}\right)^{-1} \;\mbox{cm}^2\,\mbox{s}^{-1}.
\label{eqn:conductivity}
\end{equation}
Note that $\chi$ is a \textit{thermal diffusivity}, and it depends inversely on the density.  The Spitzer \textit{conductivity} is $\kappa_{Sp} = n k_B \chi_C$ which has only the well-known $T^{5/2}$ dependence and no density dependence.  

The energy equation also includes heating ($\mathcal{H}$) and cooling ($\mathcal{L}$) terms.  The cooling function we adopt is from \citet{tn01} with the functional form  
\begin{equation}
\mathcal{L} = n_e n_p \Lambda(T),
\label{eqn:cooling1}
\end{equation}
with units of erg cm$^{-3}$ s$^{-1}$.  The temperature dependence is a fit to cooling dominated by Bremsstrahlung above 1 keV and metal lines below 1 keV with 
\begin{equation}
\Lambda(T) = \left[C_1(k_B T)^{-1.7} + C_2(k_B T)^{0.5} + C_3\right] 10^{-22},
\label{eqn:cooling2}
\end{equation}
where $C_1 = 8.6\times 10^{-3}$, $C_2 = 5.8\times 10^{-2}$, and $C_3 = 6.3 \times 10^{-2}$, for a metallicity of $Z = 0.3 Z_{\odot}$, with units of $[C_i]=\textrm{erg}\, \textrm{cm}^3\textrm{s}^{-1} $.  The majority of our runs do not include additional heating terms.  For these runs $\mathcal{H} = 0$ in equation (\ref{eqn:MHD_energy}). 

In runs with heating, we adopt a heating profile of the form
\begin{equation}
\mathcal{H} = \mathcal{H}_0 \me^{-\left(r/r_H\right)^2},
\label{eqn:heating1}
\end{equation}
where $r_H$ is the scale radius of heating and the normalization is chosen as
\begin{equation}
\mathcal{H}_0 = \frac{L_{\textrm{therm}}}{r_H^3 \pi^{3/2}},
\label{eqn:heating2}
\end{equation}
where $L_{\textrm{therm}}$ is the total thermal heating input.  This model is motivated by a simple description of AGN feedback. 
We discuss simulations with heating in more detail in \S\ref{sec:heating}.   
\subsection{Timescales}\label{sec:timescales}
We now discuss several of the key timescales in galaxy clusters in order to provide some intuition for the important physical processes (See Table \ref{tab:timescale}).  We examine these timescales for volume-averaged quantities in our fiducial atmosphere of A2199.  The generation of the fiducial atmosphere is discussed in \S\ref{sec:fidinit}.  The volume-averaged sound speed is approximately $10^8$ cm s$^{-1}$, corresponding to a sound crossing time over 50 kpc of $\tau_S \approx 45$ Myr.  A weak magnetic field of 1 nG corresponds to an Alfv\'{e}n crossing time over 50 kpc of $\tau_A \approx 10^{12}$ years. 
\begin{deluxetable}{lcc}
\tablecolumns{3}
\tablecaption{Timescales in Model T1 \label{tab:timescale}} 
\tablewidth{0pt}
\tablehead{
\colhead{Timescale} &
\colhead{Symbol} &
\colhead{Time\tablenotemark{a} (Myr)}
}
\startdata
Sound Crossing & $\tau_S$ & 45\\
Alfv\'{e}n Crossing (1$\mu$G) &$\tau_A$ & $1.1\times 10^3$\\
Conduction &$\tau_{\textrm{cond}}$ & 20 \\
HBI Growth &$\tau_{\textrm{HBI}}$ & 120\\
Cooling & $\tau_{\textrm{cool}}$ & $1.4\times 10^3$
\vspace{0.05in}
\enddata
\tablenotetext{a}{Evaluated for $L=50$ kpc as volume-averaged quantities.}
\end{deluxetable}

For a more typical magnetic field of 1 $\mu$G, the Alfv\'{e}n speed is $4.4\times 10^6$ cm s$^{-1}$, leading to a crossing time across 50 kpc of 1.1 Gyr.  We will see shortly that the Alfv\'{e}n timescale is the longest timescale in the problem.  For this cluster the central magnetic beta, the ratio of thermal pressure to magnetic pressure is $\beta_0  = 8\pi p/B^2\approx 6,600$ for $B= 1\,\mu$G.  Even for $B =10\, \mu$G, the central magnetic beta is significantly greater than unity.  Further out in the core, where the thermal pressure has dropped, the beta parameter is typically of order several hundred.

Also of interest is the heat conduction timescale.  Our model cluster has a volume-averaged thermal diffusivity of $\la \chi \ra \approx 3.8 \times 10^{31} \,\textrm{cm}^2\textrm{s}^{-1}$, which yields the scale-dependent conduction time
\begin{equation}
\tau_{\chi} = \frac{L^2}{\chi} = \left\{ \begin{array}{c c}
20 \,\textrm{Myr} & (L = 50 \,\textrm{kpc}) \\
0.80 \,\textrm{Myr} & (L = 10 \,\textrm{kpc}). \\
\end{array} \right. \end{equation}
The HBI growth time in the fast conduction limit is
\begin{equation}
\tau_{\textrm{HBI}} = \left(\frac{\dif \ln T}{\dif r}\frac{\dif \phi}{\dif r}\right)^{-1/2} \approx 126\, \textrm{Myr}.
\label{eqn:HBI-fastheat}
\end{equation}
As we will see later, the HBI has an opportunity to grow significantly compared to the average time between major mergers, roughly 5 Gyr, for a typical cluster.  The final timescale of interest is the cooling time at the center of the cluster
\be
\tau_{\textrm{cool}} = \frac{\gamma}{\gamma-1}\frac{e}{n_e n_p \Lambda(T)} \approx 1.4\,\textrm{Gyr}.
\label{eqn:cooltime}
\ee
This cooling time estimate is in general too long (by almost a factor of 2), as it does not account for the increase in the cooling rate as temperature decreases and the density and line emission increase.
Nonetheless, the cooling time is much longer than the HBI growth rate for the fastest growing, short-wavelength modes.  Thus, the HBI growth is likely to play a significant role in the thermal evolution of the cluster core.   
\subsection{Numerical Tools}\label{subsec:tools}
For our simulations we use the Athena MHD code \citep{gs08, sg08} combined with the anisotropic conduction methods of \citet{ps05} and \citet{sh07}.  In particular, we use harmonic averaging of the conductivity and the monotonized central difference limiter on transverse heat fluxes to ensure stability.  The heating, cooling, and thermal conduction are operator split and sub-cycled with respect to the MHD timestep.  The cooling simulations are implemented with a temperature floor  of $T = 0.05$ keV, below which UV lines become important, and the cooling curve fit is no longer accurate.  This temperature floor prevents the cooling catastrophe from going to completion.  

Most of the simulations in this work are done on uniform Cartesian grids of (128)$^3$.  One high-resolution run is done at (256)$^3$.  We use  modified reflecting boundary conditions for all MHD variables, in which the pressure and density are extrapolated in the ghost zones, but the other variables are reflected.  This prevents the gravitational source term from introducing a spurious acceleration at the boundary.  For the temperature boundary condition, we introduce a thermal bath at $r\ge r_{max}$ with a fixed temperature $T_{\textrm{outer}}$.  For almost all of our runs $r_{max} = 200$ kpc.   This thermal bath physically represents the massive thermal energy available in the ICM outside the cluster core.   Since we are simulating the entire cluster in a Cartesian geometry, there is no inner boundary condition at the cluster center.

To seed multiple modes of the HBI and break symmetry, we add Gaussian white noise perturbations to the initial velocity field such that the applied perturbation is everywhere a fixed fraction of the sound speed, typically $\approx 1$\%.  In the absence of the HBI, cooling, or an applied perturbation, we find that we can hold hydrostatic equilibrium to better than one part in $10^4$.  

All of our runs, unless otherwise noted, are run to a time of 9.5 Gyr, a large fraction of the age of the universe.  In a small number of runs, the simulations do not go to completion as a result of very severe cooling flows concentrating large quantities of mass into the central few zones of the grid.  These exceptions are noted.
\section{Fiducial Simulation}\label{sec:fiducial}
\subsection{Initial Conditions}\label{sec:fidinit}
To introduce the phenomenology of the HBI in cluster cores, we begin by discussing a simple fiducial calculation based on observations of the galaxy cluster Abell 2199 as discussed in \citet{zn03} and \citet{johnstone02}.  This fiducial run is identified as T1 in the table of Runs (Table \ref{tab:runs}).  Our initial conditions for the cluster core are obtained by constructing a spherically symmetric atmosphere in both hydrostatic and thermal equilibrium.  We have found that it is quite advantageous to start from thermal equilibrium.  Runs without thermal equilibrium experience thermal fronts propagating from the boundaries.  Starting in thermal equilibrium produces a much more physical initial condition. The equations for this equilibrium are
\begin{equation}
\frac{\dif  P}{\dif r} = \frac{\mu m_H P}{k_B T} \frac{\dif \phi}{\dif r},
\label{eqn:hydroeq}
\end{equation}
\begin{equation}
\frac{1}{r^2}\frac{\dif}{\dif r}\left( r^2 f_{Sp}\kappa \frac{\dif T}{\dif r}\right) = n_e n_p \Lambda(T) -\mathcal{H},
\label{eqn:thermoeq}
\end{equation}
where $f_{Sp}$ is the initial effective thermal conductivity and the heating function is neglected for our fiducial case.  Our potential is chosen to be a softened NFW potential with a dark matter density distribution given by 
\begin{equation}
\rho_{DM} = \frac{M_0/2\pi}{(r + r_c)(r + r_s)^2},
\label{eqn:NFWdens}
\end{equation}
where $M_0$ is the scale mass, $r_s$ is the scale radius, and $r_c$ is the softening radius \citep{nfw97}.  The potential softening is important for numerically maintaining a very accurate hydrostatic equilibrium.  The potential is given by
\begin{eqnarray}
\phi &=& -2GM_0 \left\{
  \frac{r_c}{(r_s - r_c)^2}\left[\ln\left(\frac{1+ r/r_c}{1+r/r_s}\right) +
        \frac{\ln\left(1 + r/r_c\right)}{r/r_c}\right] \right. \nonumber \\ 
     &-&  \left. \frac{r_s(r_s- 2r_c)}{r_c(r_s-r_c)^2}
       \frac{\ln(1+ r/r_s)}{r/r_c}\right\}.
\end{eqnarray}
The plasma is modeled as a fully ionized ideal gas with $\mu = 0.617$ and $\mu_e = 1.176$.
\begin{deluxetable*}{lccccccc}
\tablecolumns{8}
\tablecaption{Initial Properties of Nonlinear Runs \label{tab:runs}}
\tablewidth{0pt}
\tablehead{
\colhead{Run} &
\colhead{$M_0$ ($M_{\odot}$)} &
\colhead{$R_s$ (kpc)} &
\colhead{$R_c$\tablenotemark{a} (kpc)} &
\colhead{$T_i$ (keV)} &
\colhead{$T_o$ (keV)} &
\colhead{$B_0$ (G)}   &
\colhead{$K_0$ (keV cm$^{2}$)}
}
\startdata
R1 & $3.8\times 10^{14}$ & 390 & 20 & 2 & 5 & $10^{-9}$, radial & 15.5\\
T1\tablenotemark{b} & $3.8\times 10^{14}$ & 390 & 20 & 2 & 5 & $10^{-9}$, tangled & 22.4\\
T1-HB & $3.8\times 10^{14}$ & 390 & 20 & 2 & 5 & $10^{-6}$, tangled & 22.4\\
T1-256 & $3.8\times 10^{14}$ & 390 & 20 & 2 & 5 & $10^{-9}$, tangled & 22.4\\
T2 & $1.1\times 10^{15}$ & 520 & 26 & 4 & 9.5 & $10^{-9}$, tangled & 31.1\\
T3 & $3.8\times 10^{14}$ & 390 & 20 & 1 & 6 & $10^{-9}$, tangled & 5.46\\
E1 & $3.8\times 10^{14}$ & 390 & 20 & 3 & 5 & $10^{-9}$, tangled & 43.6\\
E2 & $3.8\times 10^{14}$ & 390 & 20 & 4 & 5 & $10^{-9}$, tangled & 83.1\\
E2-NC\tablenotemark{c} & $3.8\times 10^{14}$ & 390 & 20 & 4 & 5 & $10^{-9}$, tangled & 83.1\\
E2-HB & $3.8\times 10^{14}$ & 390 & 20 & 4 & 5 & $10^{-6}$, tangled & 83.1\\
E3 & $3.8\times 10^{14}$ & 390 & 20 & 4.5 & 5 & $10^{-9}$, tangled & 122.\\
E3-NC\tablenotemark{c} & $3.8\times 10^{14}$ & 390 & 20 & 4.5 & 5 & $10^{-9}$, tangled & 122.\\
H1\tablenotemark{d} & $3.8\times 10^{14}$ & 390 & 20 & 2 & 5 & $3.5\times 10^{-6}$, tangled & 14.0\\
I1\tablenotemark{e} & $3.8\times 10^{14}$ & 390 & 20 & 2 & 5 & $10^{-9}$, radial & 1.5\\
Iso1 &  $3.8\times 10^{14}$ & 390 & 20 & 4.5 & 4.5 & $10^{-9}$, tangled & 52.5\\
Iso2 &  $3.8\times 10^{14}$ & 390 & 20 & 4.5 & 4.5 & $10^{-9}$, tangled & 154.
\enddata
\tablenotetext{a}{Softening radius of NFW halo (eq. [\ref{eqn:NFWdens}])}
\tablenotetext{b}{Fiducial Run}
\tablenotetext{c}{No conduction}
\tablenotetext{d}{Simulation includes additional heating (see \S\ref{sec:heating}).}
\tablenotetext{e}{Isotropic conduction only}
\end{deluxetable*}

We solve equations (\ref{eqn:hydroeq}) and (\ref{eqn:thermoeq}) as a two point boundary value problem with the constraints of matching $T$ at both the inner ($T_i$) and outer ($T_o$) boundary.  As these are a third-order system of equations, we choose a further symmetry constraint, namely, that the heat flux vanishes at the center.  This system of equations is slightly stiff, but generally is soluble with a shooting method with a good initial guess.  We only solve these ODEs to establish our initial condition.  The subsequent evolution of the equilibrium is calculated using the full system of partial differential equations (PDEs) for MHD, equations (\ref{eqn:MHD_continuity})--(\ref{eqn:MHD_induction}).  
\begin{figure}[tbp] 
\epsscale{0.45}
\centering
\includegraphics[clip=true, scale=0.44]{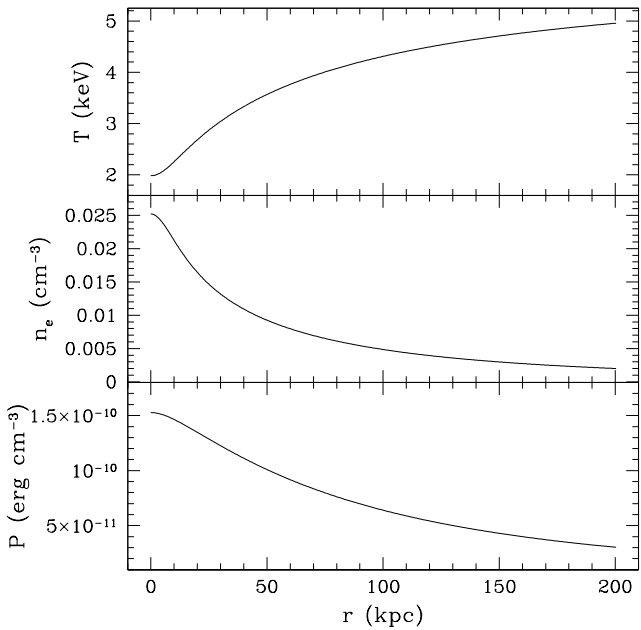}
\caption[The Fiducial Cluster Atmosphere Profiles]{The temperature, density, and pressure of our fiducial initial condition are chosen to roughly correspond to those observed in Abell 2199.  We initialize tangled magnetic fields for this simulation.}\label{fig:fidatm}
\end{figure}
 
For our fiducial model, we use a physical initial magnetic field geometry, that of tangled magnetic fields. First, in constructing the two-point boundary value for our initial conditions, we set $f_{\textrm{Sp}} = 1/3$.  Second, the initial fields are tangled with a Kolmogorov spectrum using the method outlined in detail \S4.2 of \citet{psl08}.  We initialize the fields in Fourier space as 
 \begin{equation}
\tilde{A}(k) = \tilde{A}_0\left(\frac{k}{k_\textrm{peak}}\right)^{-\alpha},
\label{eqn:A-k}
\end{equation}
where $k_{\textrm{peak}}$ is chosen as the wavenumber corresponding to 2--4 times the grid scale.  We also choose a low-$k$ cutoff corresponding to wavelengths of 1/2 to 1/4 of the domain size.  
We randomize the phase and use the Fast Fourier Transform to calculate the vector potential in real space.  We choose $\alpha = -17/6$, the appropriate $k$-space scaling for the 1D power spectrum of Kolmogorov turbulence.  Our last step in initializing the magnetic field is to difference the vector potential to calculate a manifestly divergence-free initial field.  The normalization of the magnetic field for our fiducial run is such that $\la |B|\ra=10^{-9}$ G. 

Our model cluster, Abell 2199 has an inner temperature of roughly 2 keV and a temperature of 5 keV near 200 kpc \citep{johnstone02}.  The gravitational potential is fit to an NFW profile with a scale radius of $r_s = 390$ kpc, a softening radius of $r_c = 20$ kpc, and a mass of $M_0 = 3.8\times 10^{14} \textrm{M}_{\odot}$.  Figure \ref{fig:fidatm} shows the fiducial atmosphere that results from these parameters.  The central density in our thermal equilibrium model is slightly lower than the observed value. 

\subsection{Evolution of the Fiducial Case}\label{sec:fid}
\begin{figure}[tbp!] 
\epsscale{0.45}
\centering
\includegraphics[clip=true, scale=0.44]{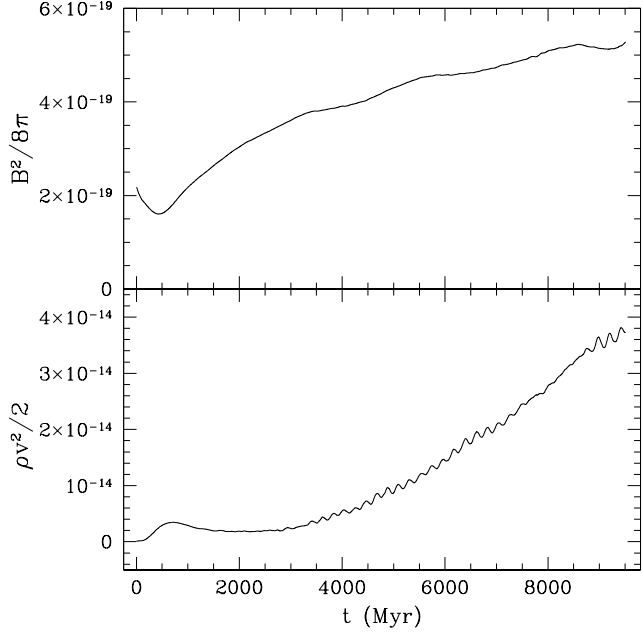}
\caption{Time evolution of the volume-averaged magnetic and kinetic energy in our fiducial run, T1.  A very weak magnetic dynamo is present.  After 2.7 Gyr, the plasma has reached the temperature floor leading to a significant inflow and magnetic field increase due to flux freezing.}\label{fig:fidtime}
\end{figure}

The evolution of our fiducial cluster model prominently illustrates the physics of the HBI and cooling in the galaxy cluster core.  This evolution is best understood through a variety of diagnostics. First, in Figure \ref{fig:fidtime}  we show the time evolution of the magnetic and kinetic energies in the cluster.  There is a very weak magnetic dynamo in the first 2 Gyr or so due to the HBI. The initial drop in magnetic energy is due to reconnection.   After $\lesssim 3$ Gyr, the core loses central pressure support, inflow begins, and the kinetic energy increases.  Correspondingly, the magnetic energy is amplified through flux-freezing leading to a maximum increase of $\Delta\la B^2\ra\approx 2.5$.  We define the magnetic energy amplification at the final simulation time, $t_f$, as
\begin{equation}
\Delta \la B^2\ra \equiv \frac{\la B^2\ra (t_f)}{\la B^2\ra (t =0)},
\label{eqn:Bamp}
\end{equation}
where the angle brackets denote a volume average.  
The kinetic energy dominates the magnetic energy at all times.  Figure \ref{fig:fidtime} shows that the HBI is not a strong source of magnetic field amplification in cluster cores. 

\begin{figure}[tbp!] 
\epsscale{0.45}
\centering
\includegraphics[clip=true, scale=0.44]{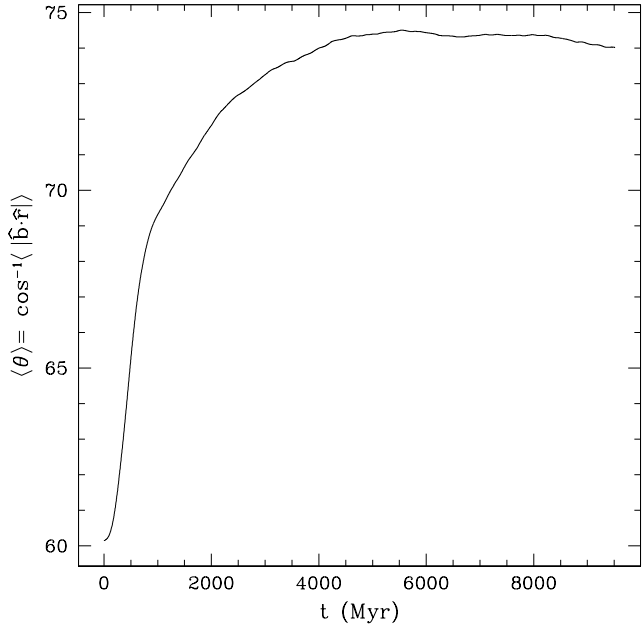}
\caption{Evolution of the volume-averaged angle of the magnetic field with respect to the radial direction in run T1.  $\theta = 0\degree$ is radial. $\theta=60\degree$ corresponds to a random magnetic field.   The magnetic field becomes significantly more azimuthal due to the HBI.}\label{fig:fidbangle}
\end{figure}

\begin{figure}[tbp] 
\epsscale{0.45}
\centering
\includegraphics[clip=true, scale=0.44]{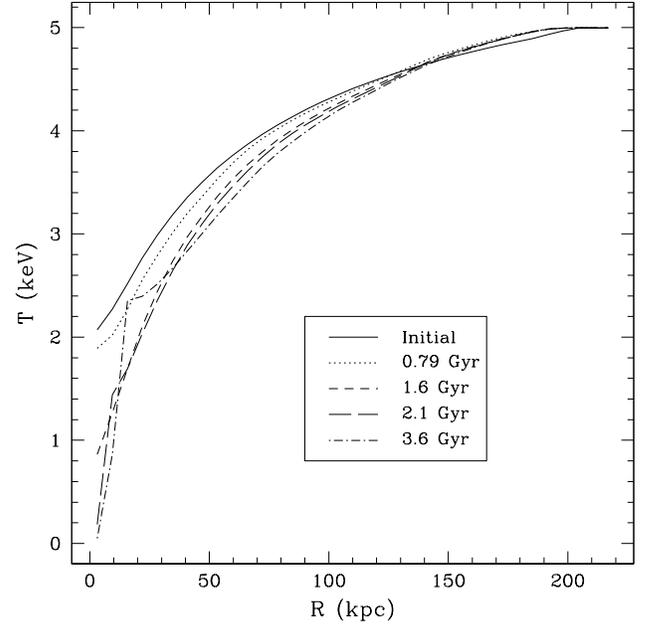}
\caption{Azimuthally-averaged radial temperature profiles in run T1.  The HBI effectively shuts off conduction leading to the cooling catastrophe that occurs around 2.7 Gyr.  The temperature is fixed at 5 keV beyond 200 kpc}\label{fig:fidtemp}
\end{figure}

The real hallmark of the HBI is the reorientation of the magnetic field geometry.  Figure \ref{fig:fidbangle} shows the evolution of the volume-averaged angle of the magnetic field with respect to the radial direction from its initial tangled state ($\theta=60\degree$) to a final saturated state of 74.6$\degree$.  The angles given here and in Table \ref{tab:satprop} are volume averages over the entire cluster.
The reorientation of the magnetic field is quite dramatic and takes place on just a few Gyr.  Concomitantly, Figure \ref{fig:fidtemp} shows the evolution of the azimuthally-averaged radial-temperature profile.  We typically bin the temperature into 5 kpc radial bins.  As the magnetic geometry evolves to be more azimuthal, the thermal conduction from outside the core begins to shut off and the temperature starts to fall. At $\sim 2.7$ Gyr, the central temperature has hit the cooling floor of $0.05$ keV---an effective proxy for the cooling catastrophe.  We do not remove gas from the grid after hitting the temperature floor, and thus, we never see a true "cooling flow.''   We define the time of the cooling catastrophe, $t_{\textrm{cc}}$, to be when the average temperature of the inner bin has reached the temperature floor.  

We can explore the magnetic field amplification in slightly more detail.
\begin{figure}[tbp] 
\epsscale{0.45}
\centering
\includegraphics[clip=true, scale=0.44]{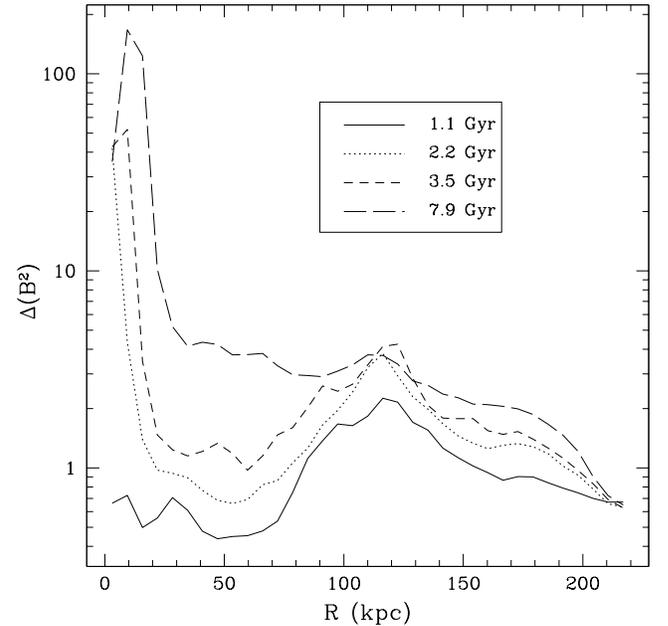}
\caption{Amplification of the magnetic energy (eq. [\ref{eqn:deltaB}]) in radial bins during run T1.  The HBI produces only a modest dynamo, which is most effective near $\sim \!100$ kpc. The central amplification of the field at late times is primarily due to flux-freezing during the cooling catastrophe.}\label{fig:fiddeltaB}
\end{figure}
We bin the magnetic field into radial bins, and then for each bin, $j$, we calculate the local amplification of the magnetic energy as
\be
\Delta (B^2)_j \equiv \frac{B_j^2 (t)}{B_j^2 (t=0)}.
\label{eqn:deltaB}
\ee
Figure \ref{fig:fiddeltaB} shows the amplification as a function of radius at several times.  The HBI amplifies the energy most efficiently in the middle of the core, just beyond 100 kpc. The high fields at late times in the central region are primarily due to flux-freezing as the density increases in the core.  

\begin{figure}[tbp!] 
\epsscale{0.45}
\centering
\includegraphics[clip=true, scale=0.44]{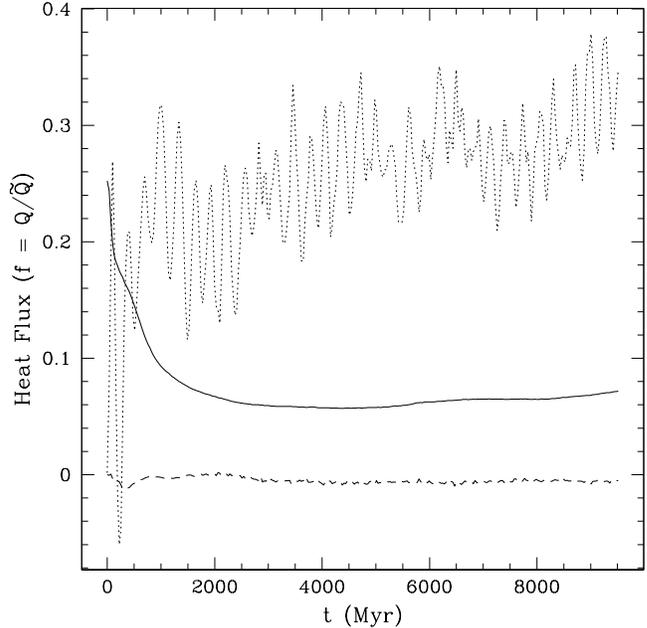}
\caption{The evolution of the components of the heat flux normalized to the instantaneous fiducial (field-free) heat flux (eq. [\ref{eqn:fidcond}]) in a shell centered at $100 \pm 40$ kpc in run T1.   The heat flux is separated into conduction (\textit{solid line}), convection (\textit{dashed line}), and mass advection (\textit{dotted line}).  The final saturated conductive heat flux has a Spitzer fraction of 7\%. }\label{fig:fidheat}
\end{figure}

Finally, Figure \ref{fig:fidheat} shows a \textit{post hoc} calculation of the heat fluxes.  See \citet{psl08} \S5.3 for a full discussion of the heat flux diagnostics.  We calculate the heat fluxes as a shell average within the radial range of $r = 100\pm 40$ kpc.  We begin by defining a fiducial heat flux to be the radial flux through the shell if the conduction were isotropic at the Spitzer value, namely:
\be
\tilde{Q}_r = - n k_B \chi_C(T, n) \frac{\dif T}{\dif r}.
\label{eqn:fidcond}
\ee
This value is the same as the heat flux with anisotropic conduction and purely radial magnetic field lines.  We calculate the conductive heat flux and normalize to the fiducial value to calculate the Spitzer fraction, or effective conductivity, defined as
\be
f_{\textrm{Sp}} \equiv Q_{\textrm{cond}}/\tilde{Q}_r,
\label{eqn:fspitz}
\ee
where $Q_{\textrm{cond}}$ is given by equation (\ref{eqn:coulombic}).
We also define a flux due to mass advection
\be
Q_{\textrm{adv}} = \frac{\gamma}{\gamma -1}k_B\left(\la n\ra\la T\ra\la v_r\ra
  + \la T\ra\la\delta n\delta v_r\ra\right),
\label{eqn:qadv}
\ee
where angle brackets denote shell averages.  The final component of the heat flux is the convective heat flux given by
\begin{eqnarray}
\label{eqn:qconv}
Q_{\textrm{conv}} &=& \frac{\gamma}{\gamma -1}k_B \left(\la n\ra\la \delta v_r \,\delta T\ra + \la v_r\ra \la\delta n\delta T\ra \right. \\ \nonumber
&+& \left. \la \delta n\delta T \delta v_r\ra\right),
\end{eqnarray}
where $\delta$ refers to the local deviation from the mean of a quantity, e.g. $\delta v_r \equiv v_r - \la v_r\ra$.  The first terms of equations (\ref{eqn:qadv}) and (\ref{eqn:qconv}) are the dominant terms.  

The initial Spitzer fraction for tangled magnetic fields is $f_{\textrm{Sp}}\approx 1/3$ due to the average over the random field geometry.  From run to run there is some variation in this initial value as there is mode power on scales larger than our averaging volume. 
As the HBI grows, the heat flux is reduced significantly, eventually saturating at $f_{\textrm{Spitz}} = 0.07$ (Fig. \ref{fig:fidheat}).  This dramatic reduction in heat flux leads to the cooling catastrophe.  As the core cools and loses pressure support, the advective heat flux increases as mass is transported inwards.  It is interesting to note that the convective heat flux is very small, especially during the HBI phase of the evolution.  

\begin{figure}[tbp] 
\epsscale{0.45}
\centering
\includegraphics[clip=true, scale=0.44]{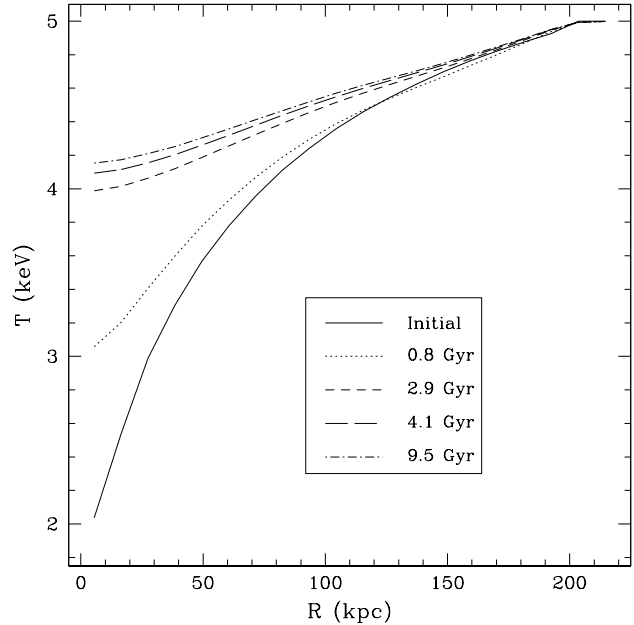}
\caption{Evolution of the azimuthally-averaged temperature profile  for run I1 with purely isotropic conduction.  Thermal instability leads to an almost isothermal state.}\label{fig:isotemp}
\end{figure}

For comparison purposes, we also run our fiducial model with purely isotropic conductivity and radial magnetic fields (run I1 in Table 1).  The HBI is not present for purely isotropic conduction.  Figure \ref{fig:isotemp} shows the evolution of the azimuthally-averaged temperature profile---the cluster reaches an almost isothermal temperature profile with no hint of the cooling catastrophe.  At fixed pressure, the thermal instability can lead to either runaway heating or runaway cooling.  This is easy to see by examining the form of the cooling term and conduction.  In the Bremsstrahlung regime, cooling scales as $T^{-3/2}$ at fixed pressure, while the conduction term scales as $T^{7/2}$.  For the case illustrated here, as the temperature is perturbed upwards, conductive heating increases much more rapidly than cooling.  Thus, the thermal runaway drives the cluster towards an almost isothermal profile.  In this simulation, there is a small amount of noise which randomizes the field a very small amount, hence $\la\theta_B\ra\ne 0$ in Table \ref{tab:satprop}.   

To summarize the fiducial case, we begin with a cluster core in hydrodynamic and thermal equilibrium.  If thermal conduction were isotropic, the central temperature would latch onto the heating branch of the thermal instability, continuing to rise.  Instead, the HBI begins to act on a 100 Myr timescale, reorienting the magnetic field geometry, and reducing the effective thermal conductivity from the outer part of the cluster.  As the cluster center becomes denser and cooler, a thermal runaway proceeds, leading to a cooling catastrophe on a timescale comparable to the initial cooling time in the core of the cluster.
\section{Variation of Parameters}\label{sec:paramvar}
In order to fully understand the behavior of the HBI in galaxy clusters, we now turn to an exploration of parameter space.  Table \ref{tab:runs} lists the various runs in which we vary the cluster properties, magnetic field strength and geometry, and the central entropy of the cluster.  The following sections describe each of these experiments.  
\begin{deluxetable}{lccccc}
\tablecolumns{6}
\tablecaption{Saturation of Nonlinear Runs \label{tab:satprop}}
\tablewidth{0pt}
\tablehead{
\colhead{Run} &
\colhead{$\Delta\la B^2\ra$} &
\colhead{max$\la\theta_B\ra$} &
\colhead{min$\fspitz$} &
\colhead{$t_{\textrm{cool}}$ (Gyr)\tablenotemark{a}} &
\colhead{$t_{\textrm{cc}}$ (Gyr)\tablenotemark{b}} 
}
\startdata
R1 & 15.0 & 65$\degree$ & 0.13 & 0.82 &4.0\\
T1 & 2.5 & 74$\degree$ & 0.07 & 1.4 & 2.7 \\
T1-HB & 1.05 & 74$\degree$ & 0.06 & 1.4 & 2.7\\
T1-256\tablenotemark{c} & 5.9 & 77\degree & 0.03 & 1.4 & 2.4\\
T2\tablenotemark{c} & 1.5 & $75\degree$ & 0.06 & 1.1 & 3.2 \\
T3 & 2.5 & $74\degree$ & 0.08 & 0.20 & 1.5 \\
E1 & 2.3 & 75$\degree$ & 0.07 & 3.0 & 3.5\\
E2 & 2.3 & 76$\degree$ & 0.05 & 5.9 & 5.7\\
E2-NC & 2.2 & 53$\degree$ & 0.48 & 5.9 & 4.0\\
E2-HB & 0.47 & 74$\degree$ & 0.08 & 5.9 & 7.0\\
E3 & 2.2 & 77$\degree$ & 0.05 & 9.3 & $>\!9.5$\\
E3-NC & 4.1 & 50.6$\degree$ & 0.55 & 9.3 & 5.9\\
H1\tablenotemark{c} &  0.30 & 72\degree & 0.09 & 1.3 & $>\!6.0$\\
I1 & 1.3 & 24\degree & 1.0 & 0.82 & none\\
Iso1 & 6.1 & 67\degree & 0.17 & 2.6 & 1.7\\
Iso2 & 1.74 & 76\degree & 0.06 & 12.8 & $>\!9.5$
\enddata
\tablenotetext{a}{Initial cooling time at the innermost radii ($\sim\! 5$ kpc).}
\tablenotetext{b}{Time of Cooling Catastrophe, when the inner gridpoint has reached the cooling floor (see \S\ref{sec:fid}).}
\tablenotetext{c}{Run time is less than 9.5 Gyr.}
\end{deluxetable}

Table \ref{tab:satprop} lists the saturation properties of these runs.  The magnetic  field amplification, $\Delta\la B^2\ra$, is given as a volume-average over the cluster.  The maximum of the magnetic field angle, $\textrm{max}\la\theta_B\ra$, is the maximum in time of the volume-averaged magnetic angle.  Likewise, $\textrm{min} f_{\textrm{Sp}}$, is the minimum in time of the shell-averaged heat flux.
\subsection{Radial Magnetic Fields}\label{sec:radial}
To assess the importance of the initial  magnetic field geometry, we consider a split monopole radial magnetic field such that $B(r) = B_c (r/r_0)^{-2}\,\textrm{sgn}(z)$.   This geometry is useful for illustrating the effects of the HBI on the equilibrium state.  We choose a mean magnetic field of 1 nG in order to minimize the effects of magnetic tension on the equilibrium state.
The resulting initial condition has a slightly higher central pressure and density than our fiducial case. In general, the atmosphere is in the rapid conduction limit on all but the largest scales.  The choice of radial magnetic fields gives conduction the best chance of thermally stabilizing the cluster.  

\begin{figure}[tbp!] 
\epsscale{0.45}
\centering
\includegraphics[clip=true, scale=0.44]{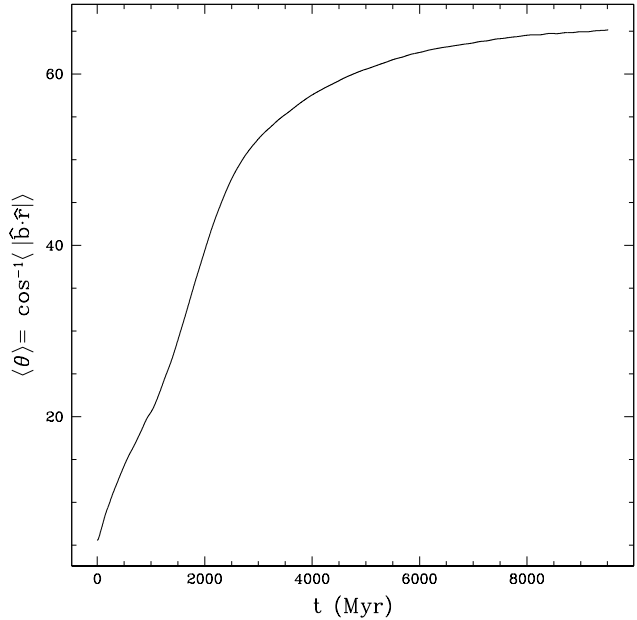}
\caption{Volume-averaged angle of the magnetic field with respect to the radial direction for run R1.  $\theta = 0\degree$ corresponds to a radial magnetic field.   The magnetic field geometry is significantly rearranged reaching a maximum of approximately $65\degree$ from radial. $\theta(t=0) \ne 0$ due to discretization errors. }\label{fig:1R-bangle}
\end{figure}

\begin{figure}[tbp] 
\epsscale{0.45}
\centering
\includegraphics[clip=true, scale=0.44]{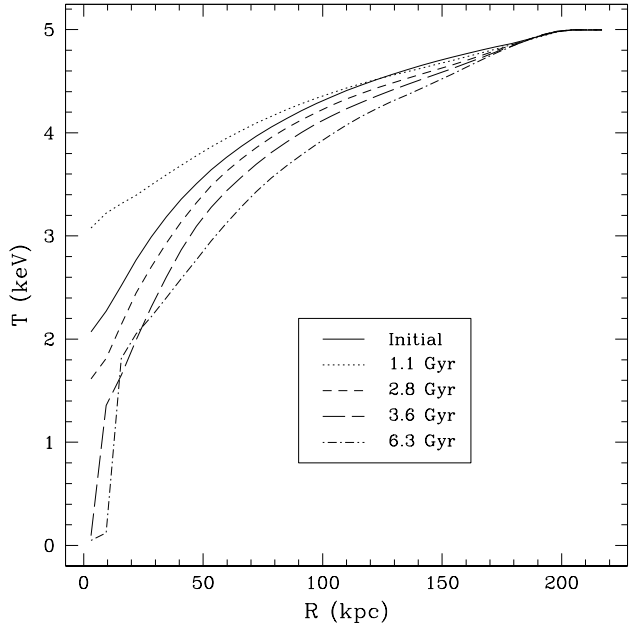}
\caption{Azimuthally-averaged radial temperature profiles in run R1.  After initially spending time on a heating branch, the HBI shuts off conduction leading to the cooling catastrophe around 3.6 Gyr. }\label{fig:1R-temp}
\end{figure}

The evolution of the HBI in the radial field simulation is quite similar to that of our fiducial case.  Initially the atmosphere is in thermal equilibrium with radial fields.  Figure \ref{fig:1R-bangle} shows the evolution of the volume-averaged magnetic field from the initially radial ($\theta \approx 0\degree$) geometry.  The HBI mode grows rapidly on a timescale of $\sim\! 100$ Myr and  reorients the magnetic field  to have $\la\theta\ra\gtrsim 60\degree$ in just over 4 Gyr.  Figure \ref{fig:1R-temp} shows the resultant evolution of the temperature profile.  The atmosphere initially latches onto a heating branch of the thermal instability, peaking at a central temperature of just over 3 keV at 1.1 Gyr.  In the absence of the HBI, it would continue to evolve to an almost isothermal state as in run I1. Instead, however, the cluster undergoes a cooling catastrophe after 4 Gyr.
\begin{figure}[tbp!] 
\epsscale{0.45}
\centering
\includegraphics[clip=true, scale=0.44]{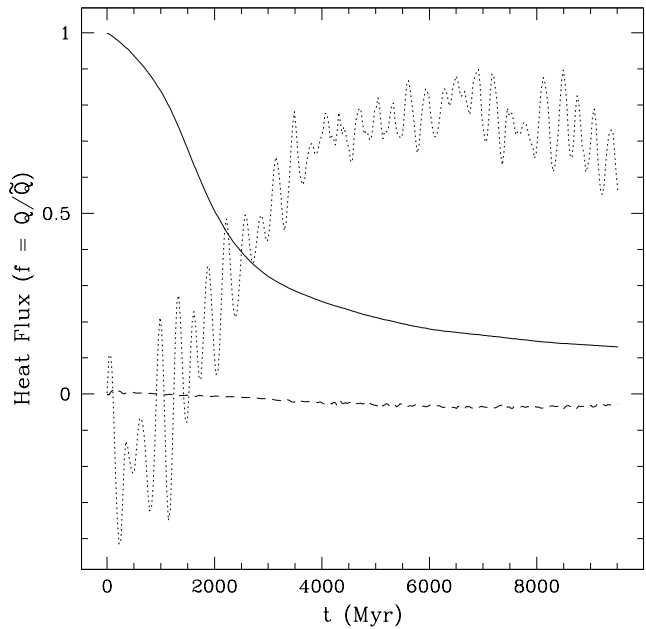}
\caption{Time evolution of the components of the heat flux normalized to the instantaneous fiducial heat flux (eq. \ref{eqn:fidcond}) in a shell centered at $100 \pm 40$ kpc for run R1.   The heat flux is separated into conduction (\textit{solid line}), convection (\textit{dashed line}), and mass advection (\textit{dotted line}).  The final saturated conductive heat flux is $\sim\! 13$\% of the field-free (Spitzer) value. }\label{fig:1R-heat}
\end{figure}
\par
\begin{figure*}[tbp!] 
\epsscale{0.45}
\centering
\includegraphics[clip=true, scale=0.65]{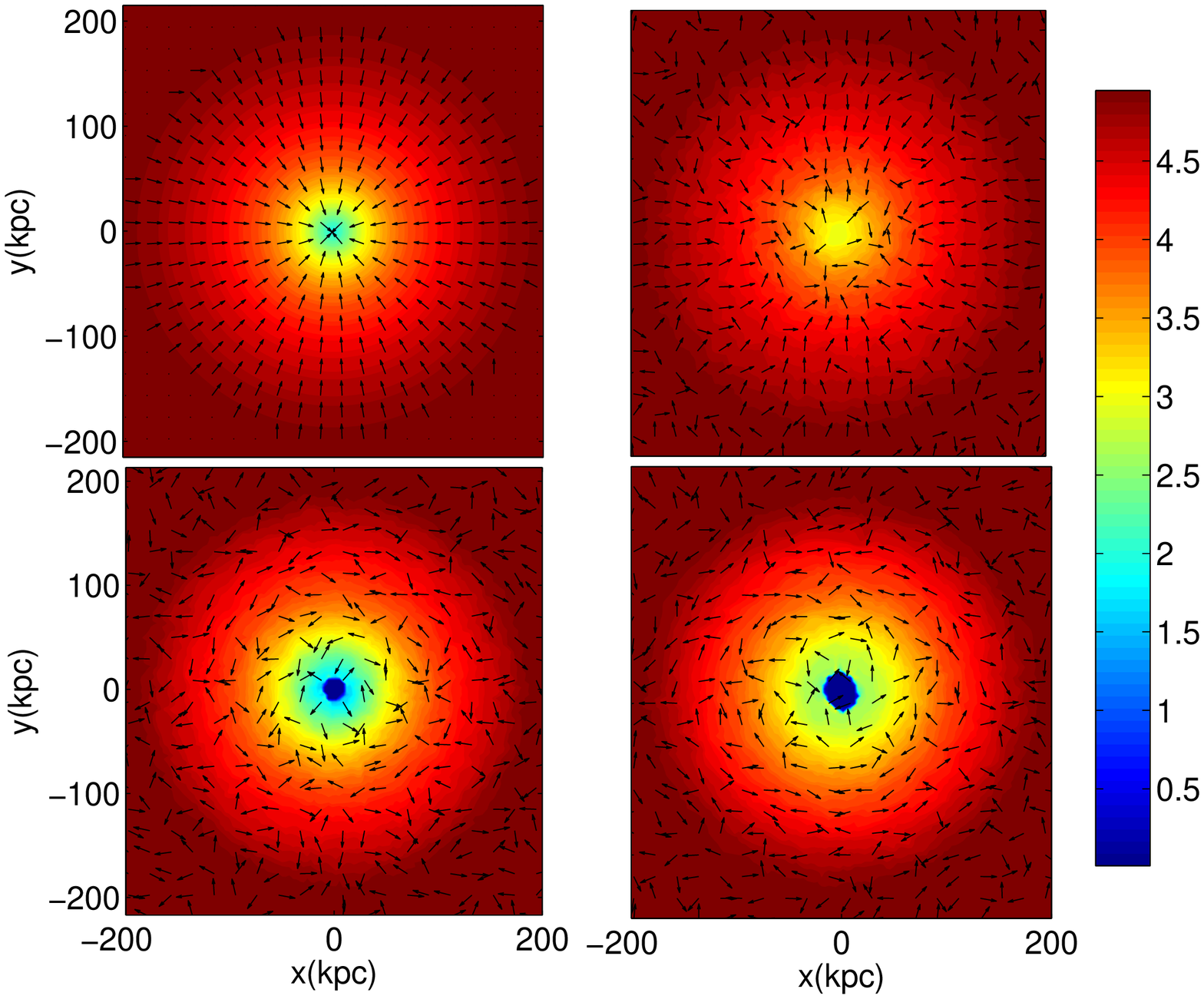}
\caption{The color scale shows the temperature in keV and the arrows represent the magnetic field unit vector in the $x$--$y$ plane for run R1 with an initially radial field.  The plots are at $t=0$, 1.6, 4.8, and 9.5 Gyr (from left to right and top to bottom).   As the magnetic field becomes more azimuthally wrapped, the cluster core reaches the cooling floor.}\label{fig:1R-visual}
\end{figure*}

 The driver of this cooling catastrophe is easily seen from Figure \ref{fig:1R-heat}, which plots the heat fluxes as a function of time at 100 kpc.  
 As the HBI rearranges the magnetic geometry, the Spitzer fraction plummets to $\fspitz\approx 0.13$.  Having reduced the contact with the thermal bath at large radii, cooling becomes dominant in the core and the central temperature starts to rapidly decrease.  This cooling drives mass inflow to small radii giving rise to the large inward advective flux at late times.  At all times, the convective heat flux is small compared to both the saturated conductive heat flux and the mixing length estimate.  Figure \ref{fig:1R-visual} shows the overall evolution of this cluster in a 2D slice taken at $z=0$.  As the magnetic field lines wrap in the azimuthal direction, the central temperature decrease is easily observed.  

The cooling catastrophe's vigor is driven by two complementary properties of the cooling curve.  First, at fixed pressure, the cooling increases as the temperature declines at the center of the cluster.  Second, below 1 keV, line emission from metals like iron and oxygen becomes increasingly important, scaling as $\mathcal{L}\propto T^{-1.7}$.  Thus, once gas has cooled below 1 keV, the cooling is much harder to reverse.  Observations show very few clusters with central temperatures below 1 keV.  
\subsection{Strong Magnetic Fields}\label{sec:strongB}
In the previous sections, we have demonstrated the evolution of the HBI in clusters for weak magnetic fields.  Under these conditions, magnetic tension is not significant.  We now consider a more realistic magnetic field of $1 \mu$G.  See \citet{ct02} for a review of cluster magnetic field measurements.  
When the tension force becomes comparable to the buoyancy time, the HBI growth is suppressed at small scales.  For example, for $B=1\,\mu$G in our fiducial cluster, magnetic tension becomes important on scales smaller than 5--20 kpc depending on the location in the cluster.

The cluster evolution of this case (labeled run T1-HB) is similar to the weaker magnetic field tangled case.  The maximum field strength we can simulate in this constant field model is limited since a modest field at the center can be dynamically important and low-$\beta$ several scale heights out from the center.  For run T1-HB we find that there is a modest amount of numerical reconnection that dissipates some of the initial magnetic energy.  A very weak dynamo leads to a maximum magnetic energy increase of only 5\%.  The maximum magnetic field angle and minimum heat flux are quite similar to the lower field case.  In addition, the cooling catastrophe occurs at almost exactly the same time.  Thus, a constant 1 $\mu$G magnetic field provides very little stabilization for our fiducial cluster.

\begin{figure}[tbp!] 
\epsscale{0.45}
\centering
\includegraphics[clip=true, scale=0.44]{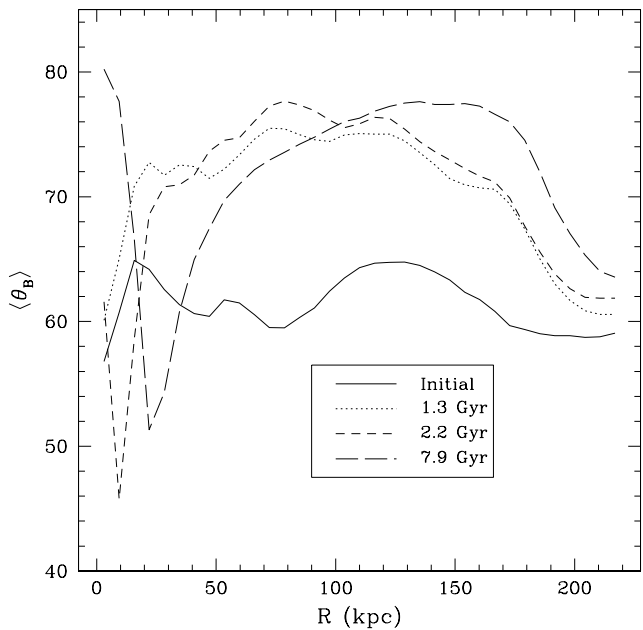}
\caption{The shell-averaged magnetic field angle versus radius for run T1-HB.  At 2.2 Gyr, radial inflow has partially reoriented the field line radially at $\sim\! 20$ kpc. }\label{fig:highB-theta}
\end{figure}

It is interesting to examine the magnetic field geometry in this simulation from a different perspective, namely, how the average magnetic field angle varies versus radius.  This is shown in Figure \ref{fig:highB-theta}. 
Initially, the angle is distributed as a random variable about 60\degree.  At 1.3 Gyr, the HBI has had somewhat more than 10 growth times to significantly increase the average angle and decrease the conductivity.  By 2.2 Gyr, radial infall from inhomogeneously cooling material has straightened the magnetic field out within 30 kpc, enhancing the radial heat flux.  This idea was suggested by \citet{br08} as a mechanism for opposing the HBI and slowing the cooling catastrophe.  Unfortunately, by the time the magnetic field has been straightened out, the plasma within 20 kpc has a mean temperature of 0.2 keV and a mean density of $\la n_e\ra \approx 1.5\,\textrm{cm}^{-3}$.  This corresponds to an increase in cooling from the initial equilibrium of $\mathcal{L}/\mathcal{L}_0\approx 350$.  Meanwhile, the average angle in this region is about 50\degree, implying that the effective conductivity has only increased a factor of 1.7 over the initial value---clearly not enough to stabilize the equilibrium. Similar behavior is found for all of our runs that reach a cooling catastrophe.
\subsection{Dependence on Cluster Parameters}\label{sec:clustvar}
We now consider the effect of varying the cluster parameters, as exemplified by runs T2 and T3 in Table \ref{tab:runs}.  We will only compare the tangled magnetic field geometries since that is the most physically relevant set-up.  Run T2 is modeled on Abell 2390, a hot, massive cluster.  This NFW halo has a larger mass and scale radius.  The core temperature rises from a central temperature of 4 keV to an outer temperature of 9.5 keV at 300 kpc.  We choose the softening radius to be approximately $1/20$ of the scale radius as we did for our fiducial case.  The dependence of our results on cluster mass and temperature is small.  The HBI growth time and central cooling time for our model of A2390 are 130 Myr and 1.1 Gyr, respectively, similar to our fiducial run.  This run evolves in a similar way to run T1, reaching the cooling catastrophe in 3.2 Gyr with similar saturated parameters.  Due to the vigor of the cooling catastrophe in this more massive case, the run does not reach 9.5 Gyr, as most of the mass has collapsed to the central few zones.  

In run T3, we examine the effect of changing the initial thermal profile but not the NFW parameters.  This run is the same as the fiducial case but the temperature now initially varies from 1 to 6 keV.  The physics of how the magnetic geometry is modified remains very similar; however, the cooling catastrophe occurs at an earlier time of 1.5 Gyr.  The primary reason for this is that the  higher initial central density and lower temperature make this cluster cool faster to the temperature floor.  Thus, we see only a very weak dependence on the cluster parameters, mostly being driven by the initial location on the cooling curve.    
\subsection{Dependence on Central Entropy}\label{sec:entropy}
Motivated by the discussion of the role of central entropy in \S\ref{sec:HBIphys}, we have undertaken a parameter study in central entropy.  The runs T1, E1, E2, and E3 form a monotonic progression of central entropy from 22.4 to 122 keV cm$^{2}$.  We have effected the entropy variation by modifying the initial central temperature while maintaining thermal equilibrium.  We fit our cluster entropy profiles with a power-law of the form of equation (\ref{eqn:entropy}).  The parameter $K_0$ is typically within 10--20\% of the central entropy.  The primary difference evident in examining Table \ref{tab:satprop} is that the time of the cooling catastrophe increases as $K_0$ increases--reaching a maximum of just over 10 Gyr for the highest entropy case with nG magnetic fields (run E3).  This phenomenon is easy to understand since the central cooling time itself increases as $K_0$ increases.  

It is interesting to compare these runs to several high-entropy simulations without conduction, runs E2-NC and E3-NC.   These runs have cooling but no thermal conduction.  First, it is clear that the central cooling time estimated by equation (\ref{eqn:cooltime}) is an overestimate compared to the actual time of the cooling catastrophe.  For example, run E3 has a predicted cooling timescale of 9.3 Gyr but in fact reaches a cooling catastrophe without conduction in 5.9 Gyr. Second, these runs without conduction show that even though the effective conductivity is reduced by the HBI, the time to a cooling catastrophe is significantly longer than in the absence of conduction.  In the case of run E3, thermal conduction delays the cooling catastrophe from 5.3 Gyr to $>\!9.5$ Gyr.

We now consider runs T1-HB and E2-HB which have higher magnetic fields of 1 $\mu$G.  In the former, the low entropy run, magnetic tension does very little to stabilize the HBI resulting in a cooling catastrophe at almost the same time as the 1 nG case.  In the latter, the high entropy run, magnetic tension plays a more significant role and increases the time of the cooling catastrophe from 5.7 Gyr (the low field case) to 7.0 Gyr (the higher field case). 
\begin{figure}[tbp] 
\epsscale{0.45}
\centering
\includegraphics[clip=true, scale=0.44]{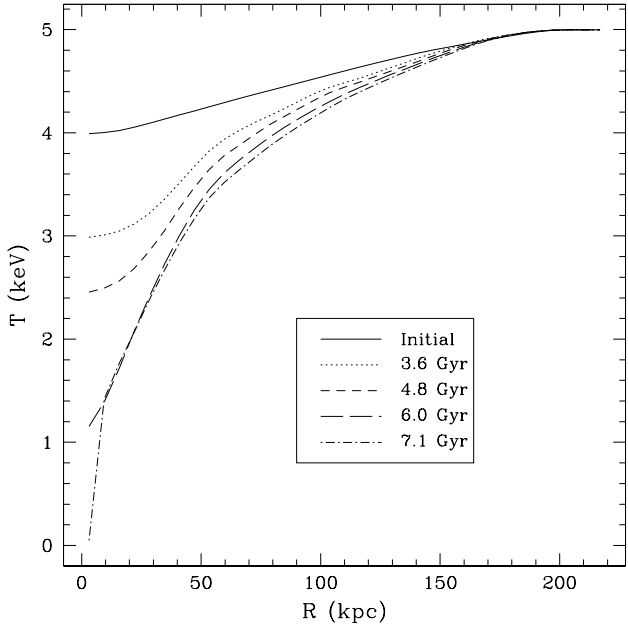}
\caption{Azimuthally-averaged temperature profiles for our high entropy (85 keV cm$^{2}$) and high magnetic field (1 $\mu$G) run (E2-HB). The HBI-driven cooling catastrophe is significantly stabilized.  The cluster does not hit the cooling floor until 7.0 Gyr.}\label{fig:E2HB-temp}
\end{figure}
Figure \ref{fig:E2HB-temp} shows the evolution of the temperature profile for run E2-HB.  The cooling is especially slow when the temperature is high.  In fact, the time to reach the cooling catastrophe, $\sim 7$ Gyr, is longer than the  typical time between major mergers for galaxy clusters $\sim\! 4$--5 Gyr \citep{cw05}. 

How does a cluster reach the high entropy states for which the cooling catastrophe can be avoided?  Cluster mergers may provide shocks that heat the cluster and boost its entropy.  In addition, strong AGN feedback may increase the cluster entropy enough to slow the cooling instability, even in the presence of the HBI.  
\subsection{Isothermal Initial Conditions}\label{sec:isothermal}
As a further exploration of the important physics in clusters, we present the results of two simulations, runs Iso1 and Iso2, that are initialized with isothermal temperature profiles at 4.5 keV.  Our usual approach of imposing thermal equilibrium (eq. 
[\ref{eqn:thermoeq}]) does not work in this case, and the central density becomes a free parameter for constructing the hydrostatic equilibrium. By definition, the initial state is not in thermal equilibrium.  The other initial parameters are the same as our fiducial case.

Run Iso1 has an initial central density of $n_{e0} = 2.5\times 10^{-2}$, which corresponds to an entropy of 52.5 keV cm$^{-1}$ and a cooling time of 2.6 Gyr.  By virtue of this short cooling time and the action of the HBI, the core experiences the cooling catastrophe at 2.1 Gyr.  By lowering the central density by a factor of 5 to $n_{e0} = 5 \times 10^{-3}$, run Iso2 has a central entropy of $\sim \!154$ keV cm$^{2}$ and a cooling time of 12.8 Gyr.  After 9.5 Gyr, this run has developed a slightly relaxed core  ($T_i\approx 4.1$ keV), but is very far from the cooling catastrophe.  

Starting from the non-equilibrium initial condition, we see very similar qualitative behavior to our equilibrium model.  Runs with short cooling times develop both the cool core and the HBI quickly, and runs with long cooling times are much more stable.  In our isothermal clusters the state is merely transient, as even clusters with long central cooling times and high entropy would eventually evolve into relaxed, cool-core clusters.
\subsection{Resolution Dependence}\label{sec:resolution}
We perform one high resolution simulation of our fiducial case, T1-256, which is a $(256)^3$ tangled field simulation.  We terminate this run at 6.3 Gyr (2/3 of the normal run time) in light of the large processor time required.  All of the qualitative results discussed thus far hold up at higher resolution.  We do see some minor differences resulting from the increased resolution:  smaller scales are now available on which the HBI can act.  The HBI both amplifies the magnetic energy slightly more and is able to reach an even larger average magnetic field angle.  The final volume-averaged magnetic field angle of $77\degree$ corresponds to an incredibly azimuthally wrapped magnetic field with a tiny effective conductivity of $f_{\textrm{Sp}}\approx 0.034$.  This precipitates the cooling catastrophe on a slightly shorter timescale.  
\subsection{Experiments with Central Heating}\label{sec:heating}
Surveys of galaxy cluster cores consistently find X-ray cavities or bubbles filled with radio emitting plasma or cosmic rays \citep[e.g.,][]{birzan04, df06}.  These structures indicative of feedback are especially prevalent in clusters with short central cooling times, roughly the same population of low entropy clusters mentioned previously.  In an effort to understand the interplay between the HBI and heating, we proceed with a preliminary analysis of heating in cluster cores.

A number of groups have proposed cosmic rays from a central AGN as a heating mechanism.  In particular, streaming cosmic rays can excite Alfv\'{e}n waves which nonlinearly Landau damp to heat the plasma \citep{lzb91}.  \citet{go08} have demonstrated in 1D models that a combination of parameterized cosmic ray feedback and conduction can prevent significant cooling for a Hubble time.  For our test problems, our heating luminosity is parameterized as in equation (\ref{eqn:heating1}) with the initial normalization set by equation (\ref{eqn:heating2}), motivated by \citet{chan05}.  These heating functions are generic and do not discriminate among cosmic ray or mechanical energy injection. 

The feedback dynamics of the cluster core is qualitatively simple.  As the core cools, the accretion rate onto the supermassive ($\sim 10^9 M_{\odot}$) black hole at the center of the cD galaxy increases.  As $\dot{M}$ increases, the feedback heating increases, slowing the cooling.  If heating becomes too effective, the accretion ceases, and a feedback loop is established.  Thus, a simple static heating model is insufficient, and we instead implement a rough time-variable version.  Namely, we sum the cooling luminosity within the heating effective radius, $r_H$, at $t=0$,
\be
\mathcal{L}_0 = \frac{\int_0^{r_H} \mathcal{L}(r) 4\pi r^2 \dif r}
{4\pi r^3/3}.
\label{eqn:l0}
\ee
We then calculate the cooling luminosity in a similar way at every timestep and scale the initial heating luminosity to the current cooling luminosity as
\be
\mathcal{H}(t) = \frac{\mathcal{L}(t)}{\mathcal{L}(t=0)}\mathcal{H}_0.
\label{eqn:hoft}
\ee
This methodology is not ideal, but given that we cannot resolve the Bondi radius on our grid, it is preferable to extrapolating the central density and temperature down to the Bondi radius; and it ensures we have an approximate feedback mechanism.  It should be noted that this heating model can be numerically unstable when the cooling instability has progressed, and large feedback heating is added within a small region.  We will improve this treatment in future work.  
\begin{figure}[tb] 
\epsscale{0.45}
\centering
\includegraphics[clip=true, scale=0.44]{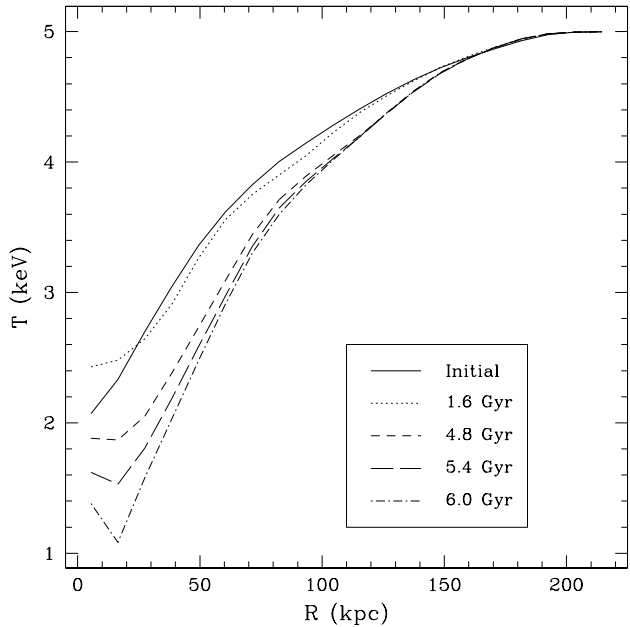}
\caption{The thermal evolution of our fiducial atmosphere with heating in the central 20 kpc, run H1.  The cooling catastrophe is pushed outwards in radius. }\label{fig:heattemp}
\end{figure}
As an example calculation, we take our fiducial atmosphere and add a total initial heating luminosity of $L_{\textrm{therm}} = 10^{43}$ erg s$^{-1}$ with a characteristic radius of $r_H = 20$ kpc.  We construct an initial thermal equilibrium such that heating, cooling, and thermal conduction all balance. This run, labeled H1, has a very interesting thermal evolution as is shown in Figure \ref{fig:heattemp}.

The HBI proceeds very slowly for this simulation since the initial average magnetic field is 3.5$\mu$G, strong enough to exert significant tension. What is especially interesting about this run is that the centrally concentrated feedback heating drives a minimum in the temperature profile at $\sim\! 20$ kpc after 5.4 Gyr of evolution. This type of profile, with a minimum slightly offset from the center, is actually observed in some clusters \citep{sand06}.  The cooling flow region seems to be simply pushed outwards from the center of the cluster, perhaps evidence that an additional volumetric heating component is needed.  The heating power at 6 Gyr has increased only modestly to a new value of $6.6\times 10^{43}$ erg s$^{-1}$.  Due to the combination of heating and the HBI the cooling catastrophe occurs much later than the initial central cooling time of 1.3 Gyr.
Unfortunately, we are not able to follow this run to completion as the sharp temperature discontinuity combined with the rapid cooling that follows leads to numerical instabilities.  While not necessarily thermally stable, this run demonstrates that cluster cores with heating and the HBI can remain stable for longer than the time between major mergers, although not necessarily a Hubble time.  
In general, it appears that the combination of magnetic fields of $\gtrsim 1 \,\mu$G (to slow the HBI) and a modest amount of AGN feedback significantly slow the cooling catastrophe, even for low-entropy clusters.  
\section{Discussion and Conclusions}\label{sec:conclusions}
The plasma in the intracluster medium of galaxy clusters is dilute and magnetized with a mean free path large compared to the gyroradius.  Under these conditions, heat transport is anisotropic along magnetic fields.  This results in the ICM being unstable to the heat-flux-driven buoyancy instability in regions where the temperature increases outward \citep{quat08}.  The cores of galaxy clusters also often have short cooling times of $\lesssim 500$ Myr.  In order to understand the thermal evolution of galaxy clusters with cooling and the HBI, we have performed three-dimensional time-dependent MHD simulations of galaxy clusters cores.

Isolated galaxy clusters evolved with magnetic fields, anisotropic conduction, and cooling share a number of common properties.  We begin with a cluster that is in both hydrostatic and thermal equilibrium. After $\sim 100$ Myr, the HBI begins to rearrange the magnetic field geometry in the cluster core;  the magnetic field saturates with an average angle between the magnetic field and the radial direction of $\sim\! 75\degree$.  Second, as the magnetic geometry is rearranged to be tangential to the temperature gradient, the magnetic field exerts a thermally insulating effect, reducing the effective radial thermal conductivity to $\lesssim 10$\% of the Spitzer value.  Finally, having reduced the thermal conduction from the outer parts of the cluster, and lacking another heat source, the core proceeds to a cooling catastrophe on a timescale comparable to the initial central cooling time.  

We have studied a number of different parameter variations and find several interesting trends.  Motivated by the observational work of \citet{voit08}, we have explored the effects of different initial entropies. For larger central cluster entropies, the time of the cooling catastrophe is delayed to more than 9.5 Gyr for our highest entropy cluster (122 keV cm$^2$).  In addition, we find that stronger magnetic fields, $\gtrsim 3.5\,\mu$G, can suppress the HBI via magnetic tension forces.  The onset of the cooling catastrophe can thus be delayed in these higher magnetic field calculations.  

We have also carried out initial calculations of the effects of heating on the ICM using a parameterized heating function in which the total heating power is proportional to the total rate of cooling in the central 20 kpc.  Despite some difficulty with numerical instabilities inherent in the method, we find that modest heating rates of $10^{43}$ erg~s$^{-1}$ can substantially delay the cooling catastrophe to $\gtrsim 5$ Gyr, longer than the time between major cluster mergers.  There is, however, some evidence that centrally concentrated heating may simply move the cooling catastrophe further out in the core (Fig. \ref{fig:heattemp}).  

Observations and simplified theoretical models both suggest that there are two distinct quasi-stable cluster states \citep{voit08, go08, cav09}.  High entropy, fairly isothermal cluster cores have long growth times for both the Field instability and the HBI.  The longer central cooling times require less conductive heating to balance cooling.  These thermal states are long-lived even in the absence of mergers.
By contrast, low entropy relaxed clusters (cool-core clusters) with short central cooling times are unstable to both the Field instability and the HBI.  As we have shown, this population of clusters cannot be stabilized by conduction alone and must have an additional feedback mechanism, plausibly the central AGN but potentially other sources.  Observations of cool core clusters with central entropies $K_0 \lesssim 30$ keV cm$^2$ show a number of feedback indicators, including H$\alpha$ emission indicative of cool gas at $10^4$ K, radio emission indicative of AGN feedback, and optical color gradients indicative of central star formation in the BCG \citep{cdv08, voit08}.  High entropy clusters, in which conduction is more important, generally show none of these feedback indicators.  Our simulations of these high entropy clusters show that they are thermally stable for cosmologically long timescales, and that conduction provides a significant stabilizing effect, e.g. runs E2 and E3 (see, Tables 2 and 3).

It may be possible for clusters to transition between these two populations.  Relaxed clusters may be promoted to high entropy clusters by significant heating, such as a major merger or an especially energetic feedback event.  Recent work shows that disruption of cool cores in a merger is possible at cosmologically early times but difficult at late times \citep{burns08}.  Alternatively, isothermal moderate entropy clusters can eventually become relaxed cool-core clusters over long timescales.  

A key task for future work is to better understand the  proposed heating mechanisms for low entropy clusters.  In particular, bubbles and jets from an AGN are far from geometrically isotropic.  Not only must the heating be locally efficient, but the heating must then be distributed by some mechanism azimuthally around the cluster core to prevent a cooling catastrophe.  The enhanced azimuthal heat transport from the HBI may play a significant role in redistributing local AGN heating throughout the cluster core.

In future work, we will examine these heating mechanisms in more detail including the relevant physics.  For the case of buoyant bubbles, there are many unanswered questions about the disruption time of the bubbles.  This shredding is governed by Rayleigh-Taylor and Kelvin-Helmholtz instabilities.  In the full Braginskii-MHD treatment, momentum is transported by ions anisotropically along magnetic field lines.  If the bubbles are indeed draped by magnetic fields, then the RT and KH instabilities will be modified by an anisotropic Braginskii viscosity. Cosmic rays may play a role in directly heating the plasma by exciting Alfv\'{e}n waves \citep{go08}.  Finally, galaxy wakes in a full cosmological context can also provide heating to the ICM \citep{conroy08}.  They may also increase the importance of thermal conduction by competing with the HBI to reorient the magnetic field.  

A key lesson of this work is that it is difficult to characterize the ICM plasma as having a single thermal conductivity  parameterized by a constant $f_{\textrm{Sp}}$.  Buoyancy instabilities such as the HBI and MTI directly modify the magnetic geometry and self-consistently evolve the system to a new state that may enhance or suppress the effective conductivity.  
In the cores of galaxy clusters, the HBI suppresses thermal conduction from the large heat reservoir at large radii.  In the absence of AGN feedback or very large magnetic fields in cluster cores, it appears that conduction alone cannot solve the cooling flow problem.  

\acknowledgements
We thank Jim Stone and Peng Oh for useful discussions.  Support for I.~J.~.P and P.~S. was provided by NASA through Chandra Postdoctoral Fellowship grants PF7-80049 and PF8-90054, respectively, awarded by the \ch X-Ray Center, which is operated by the Smithsonian Astrophysical Observatory for NASA under contract NAS8-03060.  E.~Q. was supported in part by the David and Lucile Packard Foundation, NSF-DOE Grant PHY-0812811, and NSF ATM-0752503.  Computing time was provided by the National Science Foundation through the Teragrid resources located at the National Center for Atmospheric Research and the Pittsburgh Supercomputing Center.  
\bibliography{ms}

\end{document}